\documentclass[twocolumn]{aastex631}

\usepackage{enumitem}
\usepackage{hyperref}
\usepackage{booktabs}
\usepackage{tabularx}
\usepackage[table]{xcolor}
\usepackage{amsmath}
\usepackage{mathdots}
\usepackage{multirow}
\usepackage{mathrsfs}
\usepackage{tikz}
\usepackage{float}

\usepackage{bm}

\begin{document}

% \title{Efficient parameterization of ion velocity distribution functions using Slepian functions}
\title{Strong Prevalence of Hammerhead Velocity Distributions Close to the Heliospheric Current Sheet}

\author[0000-0003-0896-7972]{Srijan Bharati Das}
\affiliation{Center for Astrophysics | Harvard \& Smithsonian, 60 Garden Street, Cambridge, MA 02138, USA}
\email{srijanbdas@alumni.princeton.edu}

\author[0000-0003-1138-652X]{Jaye L.~Verniero}
\affiliation{Code 672, NASA, Goddard Space Flight Center, Greenbelt, MD 20771, USA}

\author[0000-0002-6145-436X]{Samuel T.~Badman}
\affiliation{Center for Astrophysics | Harvard \& Smithsonian, 60 Garden Street, Cambridge, MA 02138, USA}

\author[0009-0005-9411-0378]{Robert Alexander}
\affiliation{Code 672, NASA, Goddard Space Flight Center, Greenbelt, MD 20771, USA}
\affiliation{GPHI, University of Maryland, Baltimore County, Baltimore, MD 21250, USA}

\author[0000-0003-4747-6252]{Michael Terres}
\affiliation{Center for Astrophysics | Harvard \& Smithsonian, 60 Garden Street, Cambridge, MA 02138, USA}

\author[0000-0002-5456-4771]{Federico Fraschetti}
\affiliation{Center for Astrophysics | Harvard \& Smithsonian, 60 Garden Street, Cambridge, MA 02138, USA}

\author[0000-0002-5699-090X]{Kristoff W.~Paulson}
\affiliation{Center for Astrophysics | Harvard \& Smithsonian, 60 Garden Street, Cambridge, MA 02138, USA}

\author[0000-0003-1758-6194]{Fernando Carcaboso}
\affiliation{Code 672, NASA, Goddard Space Flight Center, Greenbelt, MD 20771, USA}
\affiliation{GPHI, University of Maryland, Baltimore County, Baltimore, MD 21250, USA}

\author[0000-0001-6692-9187]{Tatiana Niembro}
\affiliation{Center for Astrophysics | Harvard \& Smithsonian, 60 Garden Street, Cambridge, MA 02138, USA}

\author[0000-0002-0396-0547]{Roberto Livi}
\affiliation{Space Sciences Laboratory, University of California, Berkeley, CA 94720-7450, USA}

\author[0000-0001-5030-6030]{Davin Larson}
\affiliation{Space Sciences Laboratory, University of California, Berkeley, CA 94720-7450, USA}

\author[0000-0003-0519-6498]{Ali Rahmati}
\affiliation{Space Sciences Laboratory, University of California, Berkeley, CA 94720-7450, USA}

\author[0000-0002-8748-2123]{Yeimy J.~Rivera}
\affiliation{Center for Astrophysics | Harvard \& Smithsonian, 60 Garden Street, Cambridge, MA 02138, USA}

\author[0000-0002-8941-3463]{Niranjana}
\affiliation{Lunar and Planetary Laboratory, University of Arizona, Tucson, AZ 85721, USA}

\author[0000-0001-6038-1923]{Kristopher G.~Klein}
\affiliation{Lunar and Planetary Laboratory, University of Arizona, Tucson, AZ 85721, USA}
\affiliation{Department of Planetary Sciences, University of Arizona, Tucson, AZ 85721, USA}

\author[0000-0002-7728-0085]{Michael L.~Stevens}
\affiliation{Center for Astrophysics | Harvard \& Smithsonian, 60 Garden Street, Cambridge, MA 02138, USA}



\begin{abstract}

The solar wind undergoes non-adiabatic heating as it travels away from the Sun. The velocity phase space distribution of non-equilibrium ions in the solar wind indicate a source of free energy that could contribute significantly to this heating. Parker Solar Probe (PSP) has observed velocity distributions containing highly anisotropic, perpendicularly diffused proton beams with a distinctly constricted gap between the core and beam populations. These distributions resemble a ``hammerhead" shape and were first reported in the fourth PSP encounter. Numerical simulations have reproduced the qualitative nature of hammerheads under certain initial conditions, but have not convincingly captured the prevalence or extreme attributes of the observed beam. This necessitates a broad study of the occurrence conditions and the associated plasma processes to better guide simulations. We statistically investigate the occurrence of these structures from 20 recent PSP encounters, and find that hammerheads dominantly occur around the Heliospheric Current Sheet (HCS). As the inclination of the HCS at PSP crossing points increases over the rising phase of the solar cycle, the occurrence of hammerheads is increasingly concentrated in narrow time periods around the HCS crossings. 
For comparison with previous work, we present statistical trends in the anisotropy of the proton beam and its connection to the density of proton beams as well as the drift speed of the beam to the core. Our study establishes a consistent occurrence pattern of hammerhead distributions around the HCS indicating hammerheads are diagnostics of energization processes associated with the HCS and its escaping wind.

\end{abstract}



% \begin{itemize}
%     \item Introduction
%     \item Methods
%     \begin{itemize}
%         \item Finding Axis of Gyrotopy [Fig 1]
%         \item Slepian Reconstruction [Fig 2, Fig 3]
%         \item In Preparation for FOV Restricted ESAs [Fig 4]
%     \end{itemize}
%     \item Final Remarks [Fig 5]
%     \item Appendix
%     \begin{itemize}
%         \item Math
%         \item Slepian on Polar Cap
%         \item Electron VDF with Spherical Harmonics: Comparison to Vinas et al. 2009.
%         \itme Ion VDF with Spherical Harmonics: Motivation for Slepian Functions
%     \end{itemize}
% \end{itemize}

\section{Introduction} \label{sec:intro}

Observations show that the solar wind expands non-adiabatically through the inner heliosphere \citep{Parker_1963, Coleman_1968, Leer_1982, Hollweg_Isenberg_2002}. However, the pathways of energy exchange (between electromagnetic fields and particles) to support this temperature profile remains an outstanding challenge. Since the plasma is weakly collisional, one of the leading theories is energization via wave-particle interactions \citep{Howes_2024, Bowen_2024_Nat, Rivera2024}. 
In this scenario, particle velocity distribution functions (VDFs) that sufficiently deviate from local thermodynamic equilibrium (LTE) generate waves that heat the plasma \citep{Marsch_2012_pVDF_Helios, verscharen2019multi}. 
In-situ measurements of solar wind ion VDFs over the last few decades have revealed a wide variety of non-LTE features such as: 
\textit{proton beams} \citep{Asbridge_1974, Goldstein_2000, Marsch_1982, Alterman_2018, Verniero_2020, Verniero_Beams_2022}, 
\textit{proton temperature anisotropy} \citep{Hundhausen_1967, Marsch_2004, Kasper_2002, Maruca_2012, Huang_2020}, \textit{alpha particle differential flow and mass-proportional temperatures}  \citep{Robbins_1970_alpha_differential_streaming,
Marsch_1982_alpha_Helios, Neugebauer_1996_alpha_Ulysses,
Steinberg_1996_alpha_WIND,
Bourouaine_2013_alpha,
Mostafavi_2022_alpha}. 
Minor ions have also been observed to be preferentially heated and accelerated compared to protons within the inner heliosphere \citep{Tracy_2015, Tracy_2016, Berger_2011, Rivera_2025}, beyond which Coulomb collisions likely constitute the dominant thermalizing effect. Investigating wave-particle interactions from a statistically large set of high-cadence observations in the inner heliosphere is necessary to understand the mechanisms responsible for the preferential heating and acceleration of ions \citep{Kasper_2017, Kasper_2019}. Parker Solar Probe \citep[PSP,][]{Fox-2016} is taking measurements closer to the Sun than any previous spacecraft, offering a unique opportunity to observe non-LTE features early in the solar wind expansion. 
The proton beams, observed by PSP, carry free energy sufficient to drive plasma instabilities, making them a prime contributor to the solar wind energy budget.

Proton VDFs are commonly partitioned into a \textit{core} component, comprising the bulk of the solar wind, as well as a suprathermal component termed the \textit{beam}. 
The presence of a two-population beam-core ion distribution in the solar wind has been regularly observed since the launch of Imp-6 in 1971 \citep{Feldman1973}. 
While past studies \citep[such as in][]{Klein_2018} have reported anisotropic proton beams, \cite{Verniero_Beams_2022} (hereafter V22) discovered the presence of a distinctly constricted gap between the perpendicularly broadened beam (relative to the magnetic field at the larger extent of their energy range) and the core, resembling a ``hammerhead" shape. 
V22 observed that these hammerhead distributions were concurrent with right-hand circularly polarized waves in the spacecraft frame. 
Perpendicularly extended proton beams resembling hammerheads have since been observed in subsequent studies in non-Alfv\'enic wind \citep{Gonzalez_2024, Malaspina_2024, Larosa_2025}. 
A few recent studies have performed simulations to analyze the generation and sustenance of these hammerhead VDFs \citep{Pezzini_2024,Shaaban_2024,Ofman_2025}.

The global structure of the heliosphere is shaped by the heliospheric current sheet (HCS), a vast, dynamic boundary that separates sectors of the Sun’s magnetic field lines of opposite polarity carried outward by the solar wind \citep{Wilcox_1965, Schulz_1973}.  As of December 2025, PSP has completed its 26th encounter; routinely cutting across the HCS \citep{Szabo_2020, fargette_HCS}. 
An unbiased investigation of properties of hammerhead distributions and their relation to global heliospheric processes requires robust statistical observations of the occurrence regions and associated electromagnetic field behavior. 
In this paper, we present an automatic detection methodology for hammerhead distributions which presents a consistent pattern of prevalent hammerhead occurrence around the HCS. 

We classify hammerheads based on the detection of a narrow \textit{neck}\footnote{Note that we adopt a slightly different nomenclature than previous studies. 
The \textit{beam} in \cite{Verniero_Beams_2022} is referred to as the \textit{neck} in our work.} connecting the core population and the perpendicularly extended portion of the beam population. 
In Sec.~\ref{sec:method} we outline the 1-dimensional (1D) convolution-based hammerhead detection method, the \textsc{Python} module \texttt{hampy}\footnote{\texttt{hampy} is hosted in this GitHub link: \href{https://github.com/srijaniiserprinceton/hampy}{https://github.com/srijaniiserprinceton/hampy}} \citep[][]{hampy}. In Sec.~\ref{sec:occurrence}, we use \texttt{hampy} around perihelia of E04 - E23 to demonstrate that hammerhead occurrences peak around the heliospheric current sheet (HCS) crossings. 
We then use the hammerheads detected by \texttt{hampy} to benchmark against the results of V22 to investigate the beam's temperature anisotropy as a function of its fractional density and its drift speed relative to the core in Sec.~\ref{sec:characterization}. There, we also present a statistical comparison of the relative densities of the core, neck and beam components of hammerheads. Finally, in Sec.~\ref{sec:discussions}, we outline a number of future questions and statistical investigations currently underway in which use hammerheads across multiple PSP encounters detected using \texttt{hampy}.

\section{Hammerhead detection methodology} 
\label{sec:method}

\begin{figure*}
    \centering
    \includegraphics[width=\linewidth]{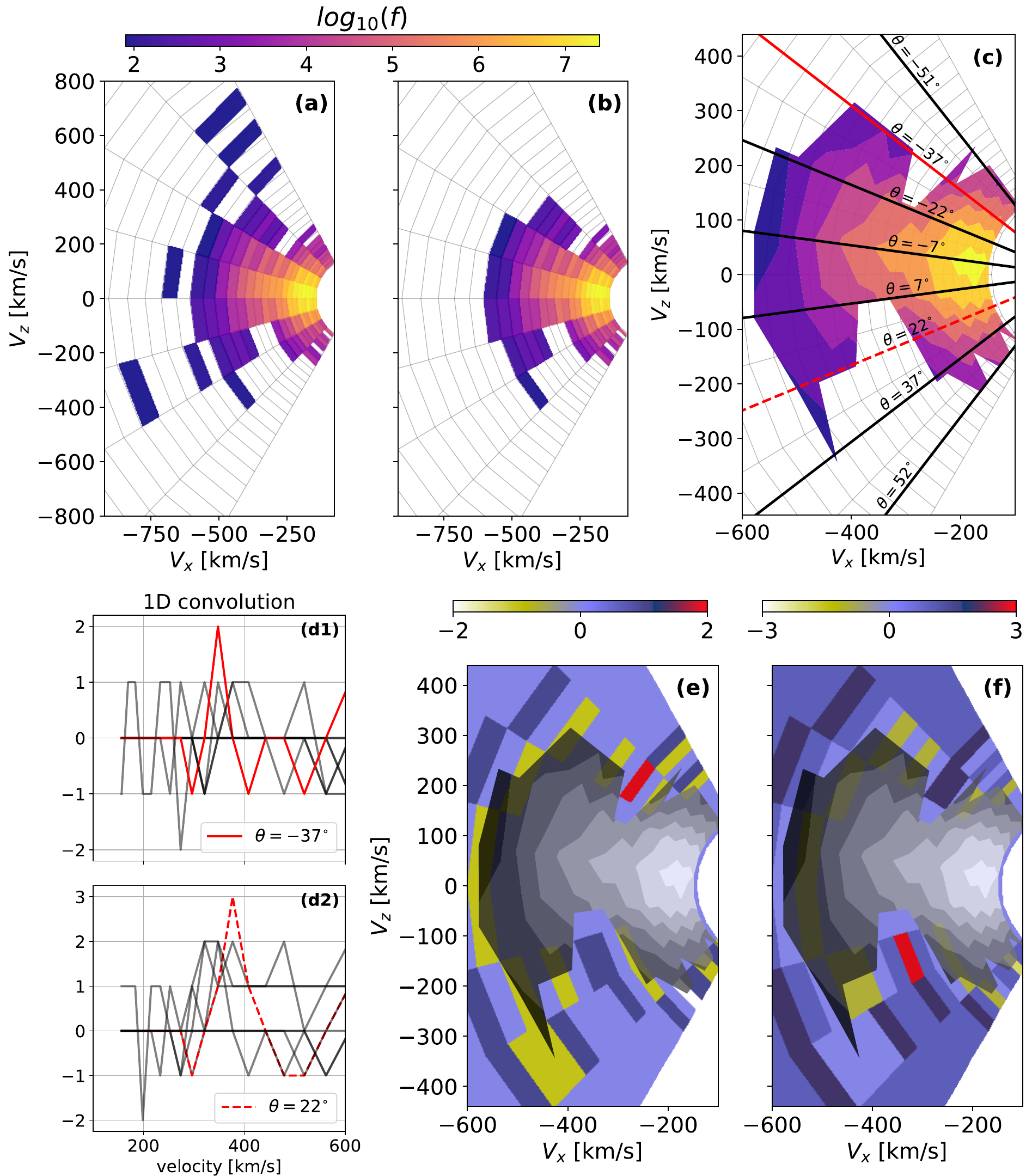}
    \caption{Schematic summary of our hammerhead detection algorithm used in \texttt{hampy}. 
    All velocity phase-space plots are presented on a $E-\theta$ plane. 
    The $V_x- V_z$ coordinates are calculated by using $\phi \sim 163.125^{\circ}$. 
    (a) $\phi$-collapsed $f(v_x, v_z)$ in the SPAN-I instrument frame before pre-processing.  
    (b) $f(v_x, v_z)$ after removing spurious count bins which do not form a contiguous distribution in the 2D $(v_x, v_z)$ space. 
    (c) Same as panel (b) but plotted using contour levels in a cell-centered convention. 
    The \textit{black} lines represent slices along different instrument elevation angles. 
    The \textit{red} lines at $\theta \sim -37^{\circ} \, \& \, 22^{\circ}$ correspond to elevations for the detected \textit{necks} for the shown hammerhead distribution. 
    The \textit{red} grids in panels (e) \& (f) show the detected necks using the corresponding 1D convolution profiles showed in panels (d1) \& (d2). 
    All VDF values are presented in log base 10.}
    \label{fig:detection_method}
\end{figure*}

We use VDF measurements from the electrostatic analyzer from SPAN-I \citep{Livi-2022} onboard PSP, part of the Solar Wind Electron Alphas and Protons \citep[SWEAP,][]{Kasper-2016} instrument suite. 
The data are publicly available on \href{https://cdaweb.gsfc.nasa.gov/}{CDAWeb} under \texttt{PSP\_SWP\_SPI\_SF00\_L2\_8DX32EX8A} with VDFs measured at a cadence ranging from 7s in earlier encounters to 1.75s in later encounters. 
We present an algorithm for a conservative detection of hammerhead VDFs directly from SPAN-I measurements. 
This involves broadly two steps --- (1) a pre-processing step to discard counts occupying non-contiguous bins in 2D velocity phase-space after collapsing the $\phi$ dimension, and (2) convolution of the pre-processed contiguous VDF with 1D gap-detection kernels along the energy dimension. 
Fig.~\ref{fig:detection_method} summarizes these steps and the mathematical details of the detection algorithm is further elaborated in the Appendix sections~\ref{sec:preprocessing} \& \ref{sec:gapfinder}. A simpler description of the method is presented below.

SPAN-I measures particle counts in a spherical polar coordinate basis $(E, \theta, \phi)$ where $E$ is the radial dimension corresponding to the energy of the recorded particle. 
The elevation $\theta$ and azimuth $\phi$ are the transverse angular dimensions (in the SPAN-I instrument frame) encoding the 3D direction of particle impingement. The primary objective for the purpose of our study is to detect a constricted neck, a gap between the core and the hammerhead beam, represented by zero counts registered by SPAN-i.
To keep our analysis simple and to avoid the FOV-limited dimension $\phi$, we perform our detection analysis on the VDF collapsed in the azimuth dimension. This results in events where hammerheads are \textit{detected} vs. \textit{not detected} by \texttt{hampy} having the same distribution as a function of anodes in $\phi$ (as illustrated in Appendix~\ref{sec: FOV_effect}). This yields a fairly robust estimate of fractional hammerhead occurrence in the context of FOV being a selection bias. The $\phi$-collapsed VDF is given by the expression
\begin{equation} \label{eqn:phi_collapsed_VDF}
    \mathcal{F}(E, \theta) = \sum_\phi f(E, \theta, \phi) \, .
\end{equation}
This summation operation is such that only those grids in $(E,\theta)$ will be flagged as ``Not a Number" (NaN) when \textit{all} the corresponding $\phi$ grids $[(E,\theta,\phi_1), ... , (E,\theta, \phi_8]]$ have zero counts. 
Conversely, for a particular $(E,\theta)$, the presence of a single finite count grid and seven empty NaN grids in $\phi$ would fail to produce an empty NaN grid in $\mathcal{F}(E,\theta)$. Since our gap finder algorithm hunts for these NaN grids in $(E,\theta)$, the above method of summation in $\phi$ naturally renders our detection algorithm conservative. Next, we discuss the filtering operation in the $(E,\theta)$ plane.

The starting point of our filtering is the $\phi$-collapsed VDF from Eqn.~(\ref{eqn:phi_collapsed_VDF}), solely in the $(E,\theta)$ plane as shown in Fig.~\ref{fig:detection_method}(a). To remove the isolated grids in $(E, \theta)$ that are not attached to a contiguous set of finite count grids, we apply a simple filtering operation (see Appendix~\ref{sec:preprocessing} for more details) which results in the non-contiguous grids being dropped as in Fig.~\ref{fig:detection_method}(b). 
The primary idea of this filtering is to remove grids that do not have at least one finite count grid neighbor in $(E, \theta)$ coordinates to either the top, bottom, left, or right. This mitigates false gap detection which would otherwise arise from spurious isolated points disconnected from the bulk of the distribution. Thereafter, a gap identification is performed via a 1D convolution with a boxcar kernel of width $N_{\mathrm{gap}}$ (see Appendix~\ref{sec:gapfinder} for more mathematical details). As shown in Fig~\ref{fig:detection_method}(c), for each elevation $\theta$, the 1D kernels are slid along $E$ to detect gaps. The 1D convolutions yield a maximal response of $N_{\mathrm{gap}}$ only when the gap size matches the box-width in the boxcar kernel. This is shown by the \textit{red} lines in Fig.~\ref{fig:detection_method} (d1) \& (d2) for $N_{\mathrm{gap}}= 2, 3$, respectively.
For each elevation angle $\theta$, our algorithm iterates through $N_{gap} \in [1,8]$ to identify the gaps. This kernel-based approach provides a robust way to identify finite contiguous gaps, ensuring that open-ended or unbounded regions are not misclassified as gaps. 
The 2-cell gap detected is shown in Fig.~\ref{fig:detection_method}(e) and the 3-cell gap detected in shown in Fig.~\ref{fig:detection_method}(f). 
We require the gaps detected on both sides of the VDF to have an overlap in energy $E$ to enforce a notion of gyrotropy. In order to avoid the typical poor-count statistics at low energy bins (which would otherwise introduce spurious necks), we also demands the neck to be outside a sphere of radius 1 Alfv\'en speed from the origin in the instrument frame. This makes our detection criteria even more conservative.

\section{Results} \label{sec:results}

We use the above described methodology encoded in our 1D convolution-based detection method, the \textsc{Python} module \texttt{hampy}, to find hammerheads across twenty PSP encounters from E04 through E23. For each encounter, we limit our analysis to $\pm 3$ days centered on the perihelion, spanning 7 days per encounter. This resulted in a total of 172,821 hammerheads detected out of 3,705,412 SPAN-I measurements.

\subsection{Occurrence of hammerheads around the Heliospheric Current Sheet} \label{sec:occurrence}

\begin{figure*}
    \centering
    \includegraphics[width=\linewidth]{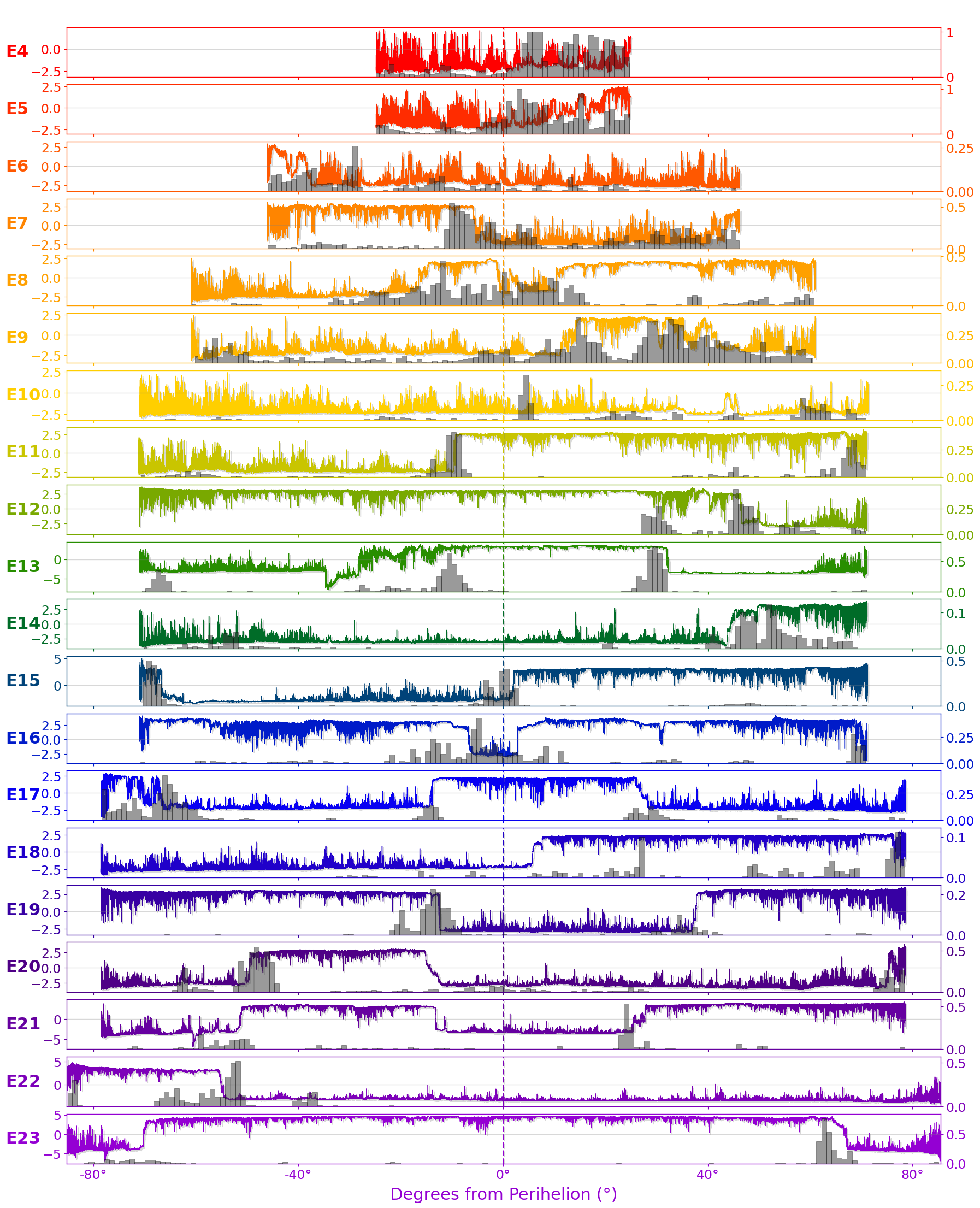}
   \caption{Figure showing (left ordinate) the scaled radial component of magnetic field $r^2 B_R(r)$ as a function of degrees from perihelion from E04 - E23. 
   The corresponding hammerhead occurrence fraction from \texttt{hampy} is plotted as gray histograms in the background (right ordinate). 
   The histograms are normalized to unit area for each encounter. 
   This figure was generated using \texttt{Plotbot}, an open-source Python package \citep{plotbot}.}
    \label{fig:hams_vs_degreesfromperihelion}
\end{figure*}

\begin{figure*}
    \centering
    \includegraphics[width=\linewidth]{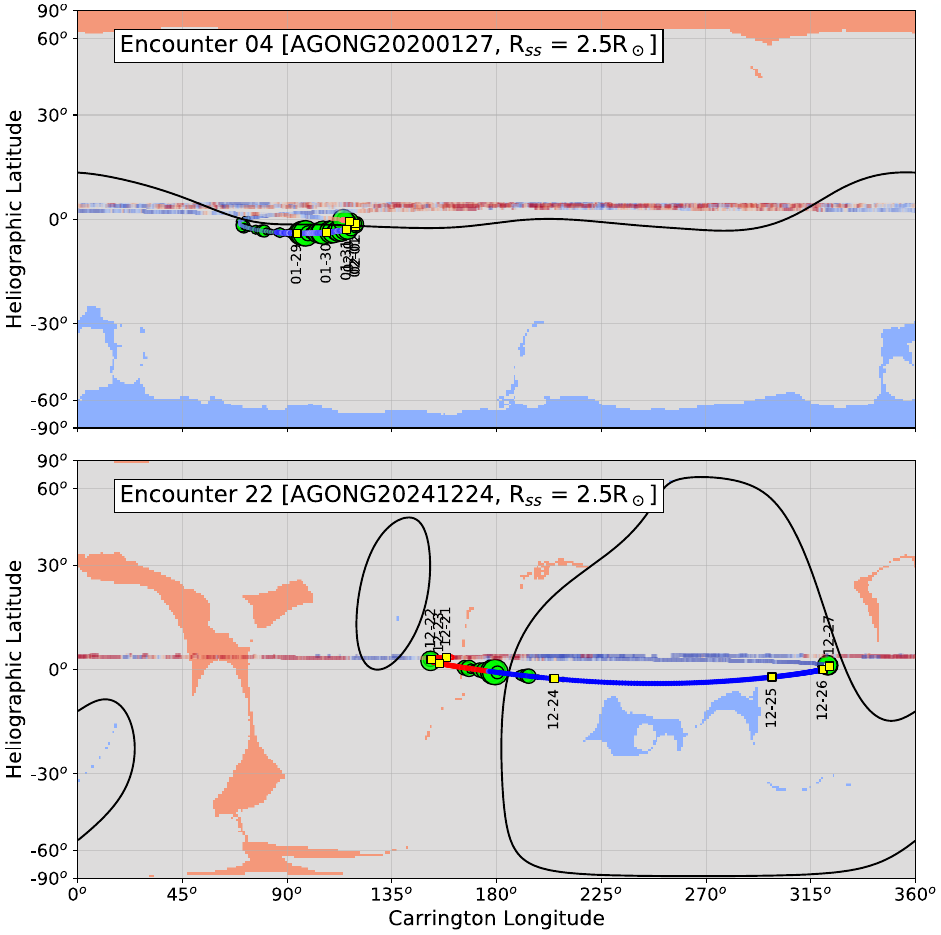}
    \caption{Heliographic projection to compare in-situ $B_R$ field sign reversals to hammerhead occurrence rate. 
    Positive in-situ $B_R$ is marked in \textit{red} and negative in-situ $B_R$ is marked in \textit{blue}. 
    PSP trajectory (back-mapped to 2.5$R_{\odot}$) is colored by $B_R$ where the perihelion is marked by higher opacity and larger markers. 
    The fainter colored parts of the trajectory fall outside the dates we consider for the encounter, where we do not use \texttt{hampy} to detect hammerheads. 
    We plot the fractional occurrence rate of hammerheads as \textit{lime-green} circles in $1^{\circ}$ bins in heliographic lat-lon. 
    The sizes of these circles are proportional to the fraction of VDFs which get flagged as hammerheads in the specific lat-lon bin. The solid \textit{black} line marks the HCS at 2.5$R_{\odot}$ using PFSS. The \textit{top} panel shows E04 while the \textit{bottom} panel shows E22. The \textit{orange} (\textit{blue}) shaded regions indicated coronal holes formed by radially outgoing (incoming) open magnetic field lines.}
    \label{fig:heliographic_maintext}
\end{figure*}

The HCS presents itself as long-lasting/sustained polarity reversals in PSP in-situ magnetic field data (most prominent in the radial component $B_R$ due to the low heliocentric distances of PSP) coincidental with reversal in electron strahl direction  \citep{fargette_HCS}. 
The left axis of Fig.~\ref{fig:hams_vs_degreesfromperihelion} displays $B_R$ (scaled by the square of distance from the Sun), as a function of (longitudinal) degrees from perihelion for E04-E23, where we color each encounter chronologically by the \textit{rainbow} palette. 

The over-plotted histogram of hammerhead occurrence fraction is shown in \textit{grey} on the right axis, computed as follows. 
The angular degree bins from perihelion are calculated using $1^{\circ}$ bins along the trajectory. 
For each bin, we record the total number of hammerheads detected $N_{\mathrm{ham}}$ and the total number of SPAN-I measurements $N_{\mathrm{total}}$. 
The height of the histogram for each bin is proportional to the occurrence fraction of hammerheads given by $N_{\mathrm{ham}}/N_{\mathrm{total}}$, where $N_{\mathrm{total}}$ is the total number of data points found in $1^{\circ}$ bin for each encounter. 
The histograms for each encounter are normalized to have unit area to aid visibility across encounters.

A visual inspection of Fig.~\ref{fig:hams_vs_degreesfromperihelion} makes plain that hammerheads are concentrated around HCS crossings. 
E11 and later encounters show that most [large-scale] $B_R$ sign reversals are associated with a localized increase in the fractional hammerhead occurrence. 
In earlier encounters E04 - E10, either (1) the HCS is not distinctly visible as a large-scale $B_R$ reversal, or (2) the hammerhead occurrences appear more spread out (compared to the tight correspondence with clear-cut HCS crossings in the later encounters). 

Figure ~\ref{fig:hams_vs_degreesfromperihelion} does not contain any information on the heliographic configuration of the HCS, apart from the large-scale $B_R$ reversal signatures measured in situ. In order to demonstrate that the hammerhead occurrence clusters around the HCS whose inclination with respect to the ecliptic rises and falls with solar activity the distribution \citep{Riley_2022}, we generate heliographic maps for each PSP encounter. We use Potential Field Source Surface (PFSS) models to estimate the global shape of the HCS during PSP encounters. An example is shown in  Fig.~\ref{fig:heliographic_maintext}, where the PSP trajectory is plotted on the heliographic coordinates along with the HCS model from a PFSS solution that has been optimized to match the magnetic polarity inversions measured in situ (see Appendix \ref{app:hcs} and Table \ref{tab:pfss} for more details). 
The PSP trajectory is colored by $r^2 B_R(r)$, where red (blue) points indicate positive (negative) $B_R$ polarity, the solid black line denotes the modeled HCS, and the size of the \textit{lime-green} circle is proportional to the hammerhead occurrence fraction $N_{\mathrm{ham}}/N_{\mathrm{total}}$ (shown as \textit{grey} histograms in Fig.~\ref{fig:hams_vs_degreesfromperihelion}, right ordinate). 
The figure contrasts an early encounter near solar minimum, E04 (top), with a later encounter near solar maximum, E22 (bottom). 
The \textit{top} panel shows that the HCS was relatively flat and grazed the PSP trajectory within a few degrees of latitude throughout E04. 
Under the hypothesis that hammerheads are spatially concentrated around the HCS, the flatness of the HCS to the orbit naturally explains the extended observations of hammerheads all along the encounter, and particularly during the second half (as also seen in the first row of Fig.~\ref{fig:hams_vs_degreesfromperihelion}). 
At the other extreme, the model HCS for E22 is nearly perpendicular to the spacecraft trajectory at each approach, and it was otherwise far removed from the PSP orbit throughout that time period.
This resulted in narrow regions of hammerhead observations around the HCS crossing and near-approaches, as shown in the \textit{bottom} panel. 
Of particular interest are the occurrences noticed around 12/22/2024 and 12/27/2024, which does not correspond to an obvious $B_R$ reversal in the in-situ data. 
The PFSS model provides support for the hypothesis here, predicting the presence of a short-lived magnetic sector boundary just beyond the eastern-most and western-most point of the orbit. 
The ``bubble" observed in the eastern-most point of the orbit, is possibly transient reconfiguration of the coronal magnetic field which may be caused by a change in the photospheric magnetic field from flux emergence or flux cancellation leading up to and during an eruption \citep{Liu2009}. 
For reference, similar plots of the hammerhead occurrence rate in the context of the PSP orbit and the PFSS current sheet position are included for all twenty encounters E04-E23 investigated in this paper  in Appendix~\ref{app:hcs}, Figs.~\ref{fig:E04-E13} \& \ref{fig:E14-E23}. Those figures likewise show that  the hammerhead observation fraction is strongly correlated with proximity to the model HCS. 
We also note hammerhead distribution observations in conjunction with several well-studied events from the PSP mission to date:
\begin{enumerate}[noitemsep, nolistsep, leftmargin=*, itemindent=0pt, parsep=0pt, topsep=0pt, partopsep=0pt]
    \item hammerheads were clustered in the quiescent regions between magnetic ``switchback'' wave patches observed across E06 \citep{Bale_2021}.
    \item the hammerhead fractional occurrence histogram peaked around the current sheet thought to be in the wake or post-eruption region of the 2022 ``Labor Day CME'' \citep{Romeo2023}.
    \item HCS crossings including embedded sunward ``jets'' (i.e. coherent flows relative to the plasma outside the HCS), were significantly less likely to be accompanied by hammerheads. These include E18, the second crossing in E20, the second crossing in E21, and the first crossing in E23. The only instance of a sunward jet that showed significant hammerhead occurrence was E14, which \cite{Phan_2024} have suggested was a complex structure comprising multiple reconnecting flux ropes. It may also be that PSP was skimming a magnetic reconnection exhaust boundary, as argued in \cite{Desai_2025}.
\end{enumerate}

\subsection{Hammerhead characteristics} \label{sec:characterization}

\begin{figure}
    \centering
    \includegraphics[width=\linewidth]{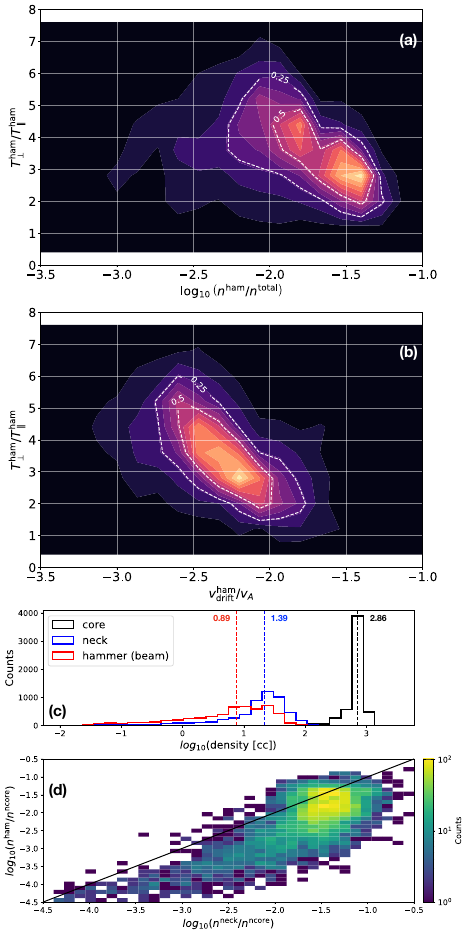}
    \caption{Characterizing hammerheads flagged by \texttt{hampy} to compare trends of temperature anisotropy of the \textit{hammer} $T_{\perp}^{\mathrm{ham}}/T_{\parallel}^{\mathrm{ham}}$ vs. (a) fractional density $n^{\mathrm{ham}}/n^{\mathrm{total}}$ and (b) normalized drift speed $v^{\mathrm{ham}}_{\mathrm{drift}}/v_{\mathrm{A}}$. 
    The 2D histograms are normalized to a maximum value of 1. 
    The \textit{white dashed} lines represent $50^{\mathrm{th}}$ and $25^{\mathrm{th}}$ quantiles. 
    The hammerheads are chosen from E04 within the radial distance bin $25-30R_{\odot}$ as a comparison against the result reported in \cite{Verniero_Beams_2022}. Panels (c) shows 1D histogram for the densities of core, neck and hammer components while panel (d) shows a 2D histogram for the fractional neck vs. the fractional hammer densities (both normalized with respect to the core). The vertical dashed \textit{black, blue} and \textit{red} lines in panel (c) show the statistical mean of the histograms for core, neck and hammer, respectively. The solid \textit{black} line in panel (d) denotes a $x=y$ line.}
    \label{fig:ham_char_V22}
\end{figure}

V22 performed a triple bi-Maxwellian fit (one Maxwellian for each core, intermediate neck, and hammerhead population) for each VDF measured by SPAN-I within a 7-hour window between 2020-01-29T16:00:00 and 2020-01-29T23:00:00. 
This interval occurred during E04 and within a 25-30$R_{\odot}$ radial distance from the Sun. 
Unlike V22, which considered every single VDF within its chosen window in the analysis, \texttt{hampy} characterizes only definitive hammerheads within a specified interval. 
Therefore, to keep sufficient statistics, we extend our analysis beyond the 7-hour window of V22 and run \texttt{hampy} within the same radial distance range 25-30$R_{\odot}$ in E04. 
We select all VDFs that are flagged as hammerheads for Fig.~\ref{fig:ham_char_V22}(a) \& (b) in order to compare with Fig. 2 of V22, with the goal of inspecting the qualitative and quantitative consistency between V22 and \textit{hampy}.
Note that we do not perform bi-Maxwellian fitting; instead, we use statistical moments. 
Once we find the \textit{neck} from our detection method (see Fig.~\ref{fig:detection_method}), we have a clear separation between the \textit{hammer} and the \textit{core}. 
We use the respective grids for the \textit{hammer} and \textit{core} to compute separate statistical moments. 
From these, we compute the total density of the VDF $n^{\mathrm{total}}$, the \textit{hammer} density $n^{\mathrm{ham}}$, the parallel drift speed of the \textit{hammer} with respect to the core $v^{\mathrm{ham}}_{\mathrm{drift}}$ and the temperature anisotropy of the \textit{hammer} population $T_{\perp}^{\mathrm{ham}}/T_{\parallel}^{\mathrm{ham}}$. 

In Fig.~\ref{fig:ham_char_V22}(a) \& (b), the peak of the histograms occurs around $T_{\perp}^{\mathrm{ham}}/T_{\parallel}^{\mathrm{ham}} \sim 2.5$, which is consistent with V22. 
Further, in V22, the temperature anisotropy for the beam was seen to asymptote to 1 (isotropy) and the fractional density $n_{\mathrm{ham}}/n_{\mathrm{total}}$ was typically around 4\%. 
Fig.~\ref{fig:ham_char_V22}(a) shows that the lower end of the $25^{\mathrm{th}}$ and the $50^{\mathrm{th}}$ quartile contours occurs around 5.6\% ($10^{-1.25}$) and 4\% ($10^{-1.4}$), respectively. 
For the dominant fraction (peak of the 2D histogram) of our flagged VDFs, the \textit{hammer} has a density of $\sim 3.1\% \, (10^{-1.5})$ of the total proton density, while V22 reported it to be around 2\%. 
Fig.~\ref{fig:ham_char_V22}(b) shows that the drift speeds of \textit{hammer} with respect to the \textit{core} ranges between $2.0 \, v_{\mathrm{A}}$ to $2.6 \, v_{\mathrm{A}}$, where $v_{\mathrm{A}}$ is the Alfv\'en speed. 
V22 reported a most probable relative drift of approximately $v_{\mathrm{drift}}^{\mathrm{ham}} = 2.5 \, v_{\mathrm{A}}$, which is consistent with the finding this method. 
We conclude from this exercise that \texttt{hampy} provides a suitable analytical framework and generalization of the phenomenon that was studied in V22 over a carefully selected period.

For the same interval, we also find from the integrated moments that $n^{\mathrm{ham}} < n^{\mathrm{neck}} \ll n^{\mathrm{core}}$, as shown in Fig.~\ref{fig:ham_char_V22}(c) \& (d). 
Fig.~\ref{fig:ham_char_V22}(c) shows that the median of the neck density is $\sim 2.4$ times the median of the hammer density. 
The median core density is consistently 1-2 orders of magnitude higher than both the neck and hammer (beam) densities. 
Fig.~\ref{fig:ham_char_V22}(d) demonstrates that in most cases, the fractional neck density $n^{\mathrm{neck}}/n^{\mathrm{core}}$ is larger than the fractional hammer density $n^{\mathrm{ham}}/n^{\mathrm{core}}$.

It is interesting to explore whether the thermal anisotropies and drifts measured here are indicative of extreme or unstable states which drive instabilities or are results of coherent waves. Prior studies, such as \cite{Klein_2021} and \cite{Martinovic_2023}, have shown that these attributes may excite ion-scale waves and convert energy rapidly, and careful measurement may shed light on the rate at which hammerheads are being created and/or dissipated in the vicinity of current sheets. SPAN-I reported moments have been used in prior studies, for instance in \cite{Liu_2025} to investigate effects of temperature anisotropy on ion-scale wave generation, or, \cite{Peng_2024} to investigate the possibility of conversion of drift kinetic energy of $\alpha$-particles to proton thermal energy due to drift instabilities, or, \cite{Amaro_2024} to study $\alpha$-to-proton temperature ratio as a function of heliographic distance. However, FOV limitations of SPAN-I onboard PSP render the moments as partial which could non-trivially affect temperature moment calculations. With this caveat in mind, we present the preliminary trends in $T_{\perp}^{\mathrm{ham}}/T_{\parallel}^{\mathrm{ham}}$ vs $n_{\mathrm{ham}}/n_{\mathrm{total}}$ and $T_{\perp}^{\mathrm{ham}}/T_{\parallel}^{\mathrm{ham}}$ vs $v_{\mathrm{drift}}^{\mathrm{ham}} / v_{\mathrm{A}}$ for all encounters E04 - E23 across nine different bins in radial distance from the Sun in Appendix~\ref{sec: Tperp_Tpara_plots} across Figs~\ref{fig:V22_replot_allEnc_fracden}~\&~\ref{fig:V22_replot_allEnc_fracvel}. 
While we do not find the same trends as reported in Fig.~\ref{fig:ham_char_V22}(a) \& (b), we hold off discussing detailed inferences for a later paper where we perform careful reconstruction of the hammerhead VDFs using recent sophisticated techniques \citep{Das_Terres_ESA} to account for possible partial FOV effects on statistical moments.

\section{Discussions and Future Work} \label{sec:discussions}

In this paper, we propose a novel technique to automatically detect hammerhead distributions in the solar wind as measured by the SPAN-I instrument onboard PSP. 
The hammerheads are defined as distributions with a distinctly constricted \textit{gap} or \textit{neck} between the core distribution and the beam (see methodology in Sec~\ref{sec:method}). 
We provide an open-source \textsc{Python} package \texttt{hampy} which relies on 1D convolution to identify these necks in VDFs and classifies them as hammerheads. 
Using twenty encounters E04 - E23 and employing \texttt{hampy} \citep{hampy}, we observe dominant occurrence of hammerhead distributions in proximity to the HCS. 
This preliminary investigation also shows hammerhead VDFs clustering around transient events such as coronal mass ejections, as well as the quiescent regions interweaving switchbacks patches. 
Future studies will investigate if the different hammerhead parameters, such as density, temperature, and drift speed could be used to distinguish between an HCS, a CME and a switchback.

Our initial study of the hammerheads using \texttt{hampy} points to several open questions. Are hammerheads the product of reconnection jets from the vicinity of the HCS or do they originate closer to the Sun? Did the reconnection jet arrive and try to mix with the core while it was already hot \citep[i.e. extended in the perpendicular direction][]{Vlad_2023}, or did the faster population arrive colder and steepened into the hammerhead shape due to \textit{in-situ} processes such as wave-particle interactions? 

Follow-up studies will incorporate the ion-scale resonant-wave properties to disentangle hammerhead origin and evolution. 
The microphysics of hammerhead distributions is intrinsically coupled to large-scale electromagnetic fluctuations, including waves, turbulence, and reconnection. Statistically relating the identified hammerheads with right-handed ion-scale waves will provide clear insights into wave-particle interactions governing these highly non-thermal distributions \citep{shankarappa2025free}.
While in-situ field observations provide a baseline, determining whether hammerheads drive or damp parallel-propagating magnetosonic waves \citep{Verniero_Beams_2022} requires rigorous dispersion analysis using tools such as ALPS \citep{Verscharen_ALPS_2018}.
Furthermore, applying advanced reconstruction techniques \citep{Das_Terres_ESA} to cast these distributions onto rectilinear grids enables the calculation of direct electromagnetic work via field-particle correlations \citep{Klein_2016} and the probing of phase space cascades \citep{Larosa_2025}. 

\section*{Acknowledgements}
    S.~B.~D, S.~T.~B, M.~T, F.~F, K.~W.~P, T.~N, R.~L, D.~L, A.~R, Y.~J.~R, K.~G.~K, N., and M.~L.~S were supported by the Parker Solar Probe mission SWEAP investigation under NASA contract NNN06AA01C. S.~B.~D and M.~T were partially supported by NASA HSR Grant 80NSSC25K7759. F.~F was partially supported by NASA award 80NSSC21K1766 (Parker Solar Probe Guest Investigator). 
    K.~G.~K and N. were partially supported by NASA Grant 80NSSC24K0724.
    We acknowledge the open policy of the data for the different missions. We acknowledge use of NASA/GSFC’s Space Physics Data Facility’s CDAWeb service. 

\appendix

\section{Hammerhead Neck Detection Methodology}

\subsection{Pre-processing to discard non-contiguous bins} \label{sec:preprocessing}

The starting point of our pre-processing is the $\phi$-collapsed VDF from Eqn.~(\ref{eqn:phi_collapsed_VDF}), solely in the $(E,\theta)$ plane as shown in Fig.~\ref{fig:detection_method}(a). To remove the isolated grids in $(E, \theta)$ that are not attached to a contiguous set of finite count grids, we apply a simple filtering operation. The primary idea of this filtering is to remove grids that do not have at least one finite count grid neighbor in $(E, \theta)$ coordinates to either the top, bottom, left, or right. 

In the pre-processing step, we remove these spurious isolated points disconnected from the bulk distribution to mitigate false gap detections. To do this, we build a 2D matrix $C_{ij}$, where $i$ corresponds to an index in $E$ and $j$ corresponds to an index in $\theta$. 
$C_{ij}$ is finite only where $\mathcal{F}(E_i, \theta_j)$ has at least one count. 
Grids which do not satisfy this criterion are populated with a ``Not a Number" entry, marked by $\mathrm{NaN}$. 
We then define a binary mask $M_{ij}$ such that

\begin{eqnarray}
M_{ij} =
\begin{cases}
1, & \text{if any of } C_{i\pm1,j},\, C_{i,j\pm1} \text{ are finite}, \\
0, & \text{otherwise.}
\end{cases}
\end{eqnarray}

The filtered field is then given by

\begin{eqnarray}
\mathcal{F}(E_i, \theta_j) =
\begin{cases}
\mathcal{F}(E_i, \theta_j), & \text{if } M_{ij} = 1 \, , \\
\mathrm{NaN}, & \text{if } M_{ij} = 0 \, .
\end{cases}
\end{eqnarray}

Our filtering effectively removes isolated finite count grids, ensuring that only adjoining bins in $(E, \theta)$ are retained for further analysis. 
This filtering operation is shown in going from Fig.~\ref{fig:detection_method}(a) to \ref{fig:detection_method}(b). 
This ensures a robust detection of the \textit{neck} and reduces false positives. Fig.~\ref{fig:detection_method}(c) is the same as Fig.~\ref{fig:detection_method}(b) but plotted using contour levels in a cell-centered convention.

% As shown in Fig.~\ref{fig:detection_method} (panels d1 and d2), the convolution yields a maximal response of $N_{gap}$ only when the gap size matches the kernel width. 
% Our algorithm iterates through $N_{gap} \in [1,8]$ to identify hammerhead populations. 
% This kernel-based approach provides a robust way to identify finite contiguous gaps, ensuring that open-ended or unbounded regions are not misclassified as gaps. 
% The 2-cell gap detected is shown in Fig.~\ref{fig:detection_method}(e) and the 3-cell gap detected in shown in Fig.~\ref{fig:detection_method}(f). 
% We require the gaps detected on both sides of the VDF to have an overlap in energy $E$ to enforce a notion of gyrotropy. 
% This makes our detection even more conservative. The final filtered VDF in $(E, \theta)$ is shown in Fig.~\ref{fig:detection_method}(b).

\subsection{Convolution using gap-finder kernels} \label{sec:gapfinder}

To identify one-dimensional gaps within the pre-processed distribution $\mathcal{F}(E,\theta)$, we used discrete 1D convolution for each elevation $\theta$ along the energy dimension --- indicated by the straight lines shown in Fig.~\ref{fig:detection_method}(c). As shown in the course of this section, the presence of such symmetric gaps is how we quantify the presence of a neck.
A family of boxcar-shaped kernels are designed to detect contiguous ``gap cells" enclosed by ``non-gap cells" on both sides as we scan along $E$ for a given value of elevation $\theta$. Below we provide a mathematical description for the gap finder kernels.

Let \( M_i(E) \) denote the binary mask along direction \(E\) 
for elevation grid index \(i\) to flag zero-count grids in $(E,\theta)$.
\[
M_i(E_j) =
\begin{cases}
1, & \text{if } F(E_j,\theta_i) \text{ is NaN}, \\[4pt]
0, & \text{if } F(E_j,\theta_i) \text{ is finite.}
\end{cases}
\]
For each possible gap length \( N_{\mathrm{gap}} \in \{1, 2, \ldots, N_{\mathrm{gap}}^{\max}\} \),
we define a one-dimensional \textit{gap-detector} kernel
\[
h_{N_{\mathrm{gap}}}(E_j) = 
\begin{cases}
-1, & j = 0, \\[4pt]
\hspace{0.25cm} 1, & j = 1, 2, \ldots, N_{\mathrm{gap}}, \\[4pt]
-1, & j = N_{\mathrm{gap}} + 1, \\[4pt]
\hspace{0.25cm} 0, & \text{otherwise.}
\end{cases}
\]
We define $N_{\mathrm{gap}}^{\mathrm{max}}=8$ since (A) we never detected neck lengths to be larger than 8 cells wide, and, (B) we prescribed the maximum gap in grids to not exceed the maximum number of grids in elevation $\theta$. The 1D convolution between the mask at a particular elevation $\theta_i$ and the gap-detector kernel is computed as
\begin{eqnarray}
C_{\theta_i}(E_j; N_{\mathrm{gap}}) &=& M_i(E_j) \, * \,  h_{N_{\mathrm{gap}}}(E_j) \, , \nonumber \\ 
C_{\theta_i}(E_j; N_{\mathrm{gap}}) &=& \sum_{k=E_0}^{E_{31}} M_i(E_k)\, h_{N_{\mathrm{gap}}}(E_j - E_k) \, .
\end{eqnarray}

For the pre-processed VDF in Fig.~\ref{fig:detection_method}(b), $C_{\theta=-37^{\circ}}(E_j; 2)$ is shown in Fig.~\ref{fig:detection_method}(d1) and $C_{\theta=22^{\circ}}(E_j; 3)$ is shown in Fig.~\ref{fig:detection_method}(d2). 
A valid gap of length \( N_{\mathrm{gap}} \) is detected when at a given elevation $\theta_i$ and centered at energy $E_j$ we get
\[
C_{\theta_i}(E_j; N_{\mathrm{gap}}) = N_{\mathrm{gap}},
\]
which corresponds to a region where the \(N_{\mathrm{gap}}\) consecutive 
cells are marked as gaps. 
This is shown by the \textit{red} lines in Figs.~\ref{fig:detection_method}(d1) \& (d2). 
Both the left and right boundaries of kernel $h_{N_{\mathrm{gap}}}$ contribute negative terms from the 
flanking \(-1\) entries in the kernel, making sure the gap detected for kernel $h_{N_{\mathrm{gap}}}$ is precisely $N_{\mathrm{gap}}$ in length. 
This kernel-based approach provides a robust way to identify finite contiguous gaps, ensuring that open-ended or unbounded regions are not misclassified as gaps. 
The 2-cell gap detected is shown in Fig.~\ref{fig:detection_method}(e) and the 3-cell gap detected in shown in Fig.~\ref{fig:detection_method}(f). 
We require the gaps detected on both sides of the VDF to have an overlap in energy $E$ to enforce a notion of gyrotropy. This further makes our detection algorithm more conservative. In order to prevent being susceptible to low energy count statistics of the SPAN-i ESA, we also require such symmetric gaps to occur beyond $v_A$ from the origin in the instrument frame. Detections which satisfy all these criteria are flagged as hammerheads and used in the downstream processing presented in the results of this paper.

\section{Investigating FOV effects on hammerhead detection using \texttt{hampy}} \label{sec: FOV_effect}

\begin{figure}[H]
    \centering
    \includegraphics[width=0.8\linewidth]{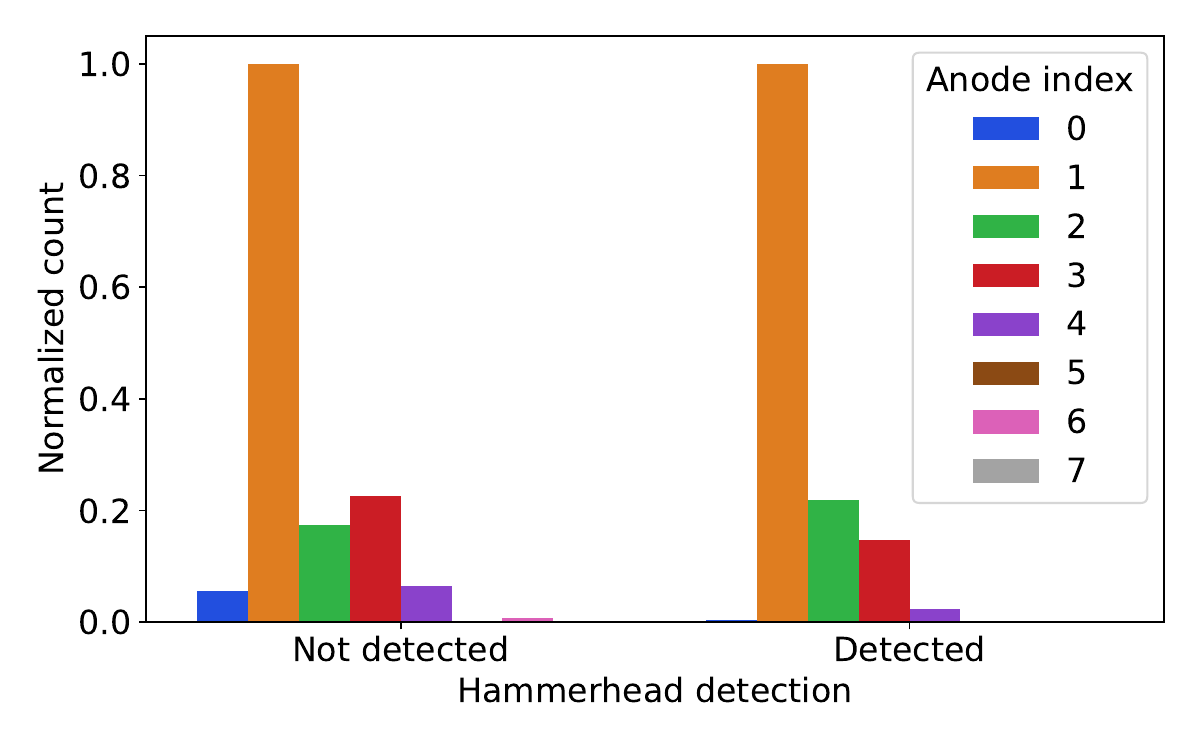}
    \caption{Histogram as a function of SPAN-i anode index for cases where \texttt{hampy} did not flag a VDF as hammerhead vs. when it was flagged as a hammerhead. The anode index is colored according to the legend shown. The counts are normalized to have a maximum value of 1 for both the `Not detected' and `Detected' cases. We considered measurements during perihelia across E04 - E23 in this figure.}
    \label{fig:barplot_allEnc}
\end{figure}

Since the SPAN-I instrument has a heat-shield obstruction, the natural question is if the hammerhead detection is potentially biased by FOV effects. 
In order to address this, we bin all events during $\pm 3$ days around perihelia of E04 - E23 based on which SPAN-I anode contains the peak of the VDF. 
We divide the samples into those that are labeled to be hammerheads by \texttt{hampy} and those that are not. 
The resultant histogram is presented in Fig.~\ref{fig:barplot_allEnc}. 
Note that due to heat-shield aberrations the closest anode (here labeled as `0') has a subdued occurrence of VDF peaks. 
However, the rest of the distribution across anode index is similar for both the `Detected' and the `Not detected' cases. 
This shows that hammerhead detection is not largely biased by selection effects arising from limited FOV coverage. 
Since \texttt{hampy} collapses the VDF in anodes and only uses the energy-elevation $(E-\theta)$ plane to detect the gaps, it is not expected that limited coverage in the anode $\phi$ plane would heavily bias the detection algorithm.

\section{Encounter-wise hammerhead proximity to HCS}\label{app:hcs}

\begin{table}[H]
\centering
\begin{tabular}{|l|l|l|l|}
\hline
\parbox[t]{0.10\textwidth}{\textbf{Encounter}} &
\parbox[t]{0.20\textwidth}{\textbf{HCS model}} &
\parbox[t]{0.20\textwidth}{\textbf{Literature Reference}} &
\parbox[t]{0.40\textwidth}{\textbf{Comments}} \\ \hline

\parbox[t]{0.10\textwidth}{E04} &
\parbox[t]{0.20\textwidth}{2020/01/27, 2.5 $R_\odot$} &
\parbox[t]{0.20\textwidth}{\citet{Liewer2025}} &
\parbox[t]{0.40\textwidth}{Flat HCS/solar min; IPs 2020/01/30, 21:50, uncertain RH.} \\ \hline

\parbox[t]{0.10\textwidth}{E05} &
\parbox[t]{0.20\textwidth}{2020/06/07, 2.5 $R_\odot$} &
\parbox[t]{0.20\textwidth}{\citet{Liewer2025}} &
\parbox[t]{0.40\textwidth}{Flat HCS/solar min; IPs 2020/06/08, 02:47} \\ \hline

\parbox[t]{0.10\textwidth}{E06} &
\parbox[t]{0.20\textwidth}{2020/09/27, 2.5 $R_\odot$} &
\parbox[t]{0.20\textwidth}{Unchanged from perihelion/canonical $R_{SS}$} &
\parbox[t]{0.40\textwidth}{Hams in quiescent periods between SB patches; IPs 2020/09/25, 03:39, uncertain RH.} \\ \hline

\parbox[t]{0.10\textwidth}{E07} &
\parbox[t]{0.20\textwidth}{2021/01/17, 2.5 $R_\odot$} &
\parbox[t]{0.20\textwidth}{Unchanged from perihelion/canonical $R_{SS}$} &
\parbox[t]{0.40\textwidth}{---} \\ \hline

\parbox[t]{0.10\textwidth}{E08} &
\parbox[t]{0.20\textwidth}{2021/05/01, 2.5 $R_\odot$} &
\parbox[t]{0.20\textwidth}{\citet{Kasper2021, 2023FrASS..1091294N}} &
\parbox[t]{0.40\textwidth}{CME associated with a filament eruption near an HCS crossing.} \\ \hline

\parbox[t]{0.10\textwidth}{E09} &
\parbox[t]{0.20\textwidth}{2021/08/13, 1.8 $R_\odot$} &
\parbox[t]{0.20\textwidth}{Newly optimized in this work} &
\parbox[t]{0.40\textwidth}{Low source surface height needed} \\ \hline

\parbox[t]{0.10\textwidth}{E10} &
\parbox[t]{0.20\textwidth}{2021/11/15, 2.5 $R_\odot$} &
\parbox[t]{0.20\textwidth}{\citet{Badman2023}} &
\parbox[t]{0.40\textwidth}{Unipolar perihelion, CME or close HCS approach} \\ \hline

\parbox[t]{0.10\textwidth}{E11} &
\parbox[t]{0.20\textwidth}{2022/02/24, 2.5 $R_\odot$} &
\parbox[t]{0.20\textwidth}{\citet{Ervin2024a, Rivera2024, Rivera2025}} &
\parbox[t]{0.40\textwidth}{First near vertical HCS crossing;} \\ \hline

\parbox[t]{0.10\textwidth}{E12} &
\parbox[t]{0.20\textwidth}{2022/05/30, 2.5 $R_\odot$} &
\parbox[t]{0.20\textwidth}{Newly optimized in this work} &
\parbox[t]{0.40\textwidth}{---} \\ \hline

\parbox[t]{0.10\textwidth}{E13} &
\parbox[t]{0.20\textwidth}{2022/09/15, 2.5 $R_\odot$} &
\parbox[t]{0.20\textwidth}{\citet{Romeo2023, 2025NatAs...9.1444P}} &
\parbox[t]{0.40\textwidth}{Large CME deforms HCS and produce CME-associated current sheets.} \\ \hline

\parbox[t]{0.10\textwidth}{E14} &
\parbox[t]{0.20\textwidth}{2022/12/11, 2.5 $R_\odot$} &
\parbox[t]{0.20\textwidth}{Unchanged from perihelion/canonical $R_{SS}$} &
\parbox[t]{0.40\textwidth}{Proximity to disconnected ``bubble''} \\ \hline

\parbox[t]{0.10\textwidth}{E15} &
\parbox[t]{0.20\textwidth}{2023/03/16, 2.5 $R_\odot$} &
\parbox[t]{0.20\textwidth}{\citet{Ervin2024b}} &
\parbox[t]{0.40\textwidth}{CME/sub-sonic interval inbound} \\ \hline

\parbox[t]{0.10\textwidth}{E16} &
\parbox[t]{0.20\textwidth}{2023/06/26, 2.4 $R_\odot$} &
\parbox[t]{0.20\textwidth}{Newly optimized in this work} &
\parbox[t]{0.40\textwidth}{Disconnected ``bubble'' at perihelion; wave-ham interaction investigated in \cite{Shi_2024}} \\ \hline

\parbox[t]{0.10\textwidth}{E17} &
\parbox[t]{0.20\textwidth}{2023/09/27, 2.5 $R_\odot$} &
\parbox[t]{0.20\textwidth}{Unchanged from perihelion/canonical $R_{SS}$} &
\parbox[t]{0.40\textwidth}{---} \\ \hline

\parbox[t]{0.10\textwidth}{E18} &
\parbox[t]{0.20\textwidth}{2023/12/29, 2.5 $R_\odot$} &
\parbox[t]{0.20\textwidth}{Unchanged from perihelion/canonical $R_{SS}$} &
\parbox[t]{0.40\textwidth}{Sunward exhaust and low ham occurrence.} \\ \hline

\parbox[t]{0.10\textwidth}{E19} &
\parbox[t]{0.20\textwidth}{2024/03/30, 2.5 $R_\odot$} &
\parbox[t]{0.20\textwidth}{Unchanged from perihelion/canonical $R_{SS}$} &
\parbox[t]{0.40\textwidth}{---} \\ \hline

\parbox[t]{0.10\textwidth}{E20} &
\parbox[t]{0.20\textwidth}{2024/07/05, 3.0 $R_\odot$} &
\parbox[t]{0.20\textwidth}{Newly optimized in this work} &
\parbox[t]{0.40\textwidth}{Second crossing has sunward exhaust and low ham occurrence.} \\ \hline

\parbox[t]{0.10\textwidth}{E21} &
\parbox[t]{0.20\textwidth}{2024/09/30, 2.5 $R_\odot$} &
\parbox[t]{0.20\textwidth}{Unchanged from perihelion/canonical $R_{SS}$} &
\parbox[t]{0.40\textwidth}{Second crossing has sunward exhaust and low ham occurrence.} \\ \hline

\parbox[t]{0.10\textwidth}{E22} &
\parbox[t]{0.20\textwidth}{2024/12/24, 2.5 $R_\odot$} &
\parbox[t]{0.20\textwidth}{Unchanged from perihelion/canonical $R_{SS}$} &
\parbox[t]{0.40\textwidth}{Proximity to disconnected ``bubble'' on inbound} \\ \hline

\parbox[t]{0.10\textwidth}{E23} &
\parbox[t]{0.20\textwidth}{2025/03/22, 2.5 $R_\odot$} &
\parbox[t]{0.20\textwidth}{Unchanged from perihelion/canonical $R_{SS}$} &
\parbox[t]{0.40\textwidth}{First crossing has sunward exhaust and low ham occurrence.} \\ \hline

\end{tabular}
\caption{Listing of PFSS parameters and comments on hammerhead VDF association for PSP encounters E04 - E23. The ``Comments'' column is written in reference to Figs.~\ref{fig:E04-E13} \& \ref{fig:E14-E23}. Interplanetary shocks (IPs) and Rankine-Hugoniot (RH) analysis thereof are reported in \cite{Kruparova_2025} database.}
\label{tab:pfss}
\end{table}

\begin{figure}
    \centering
    \includegraphics[width=\linewidth]{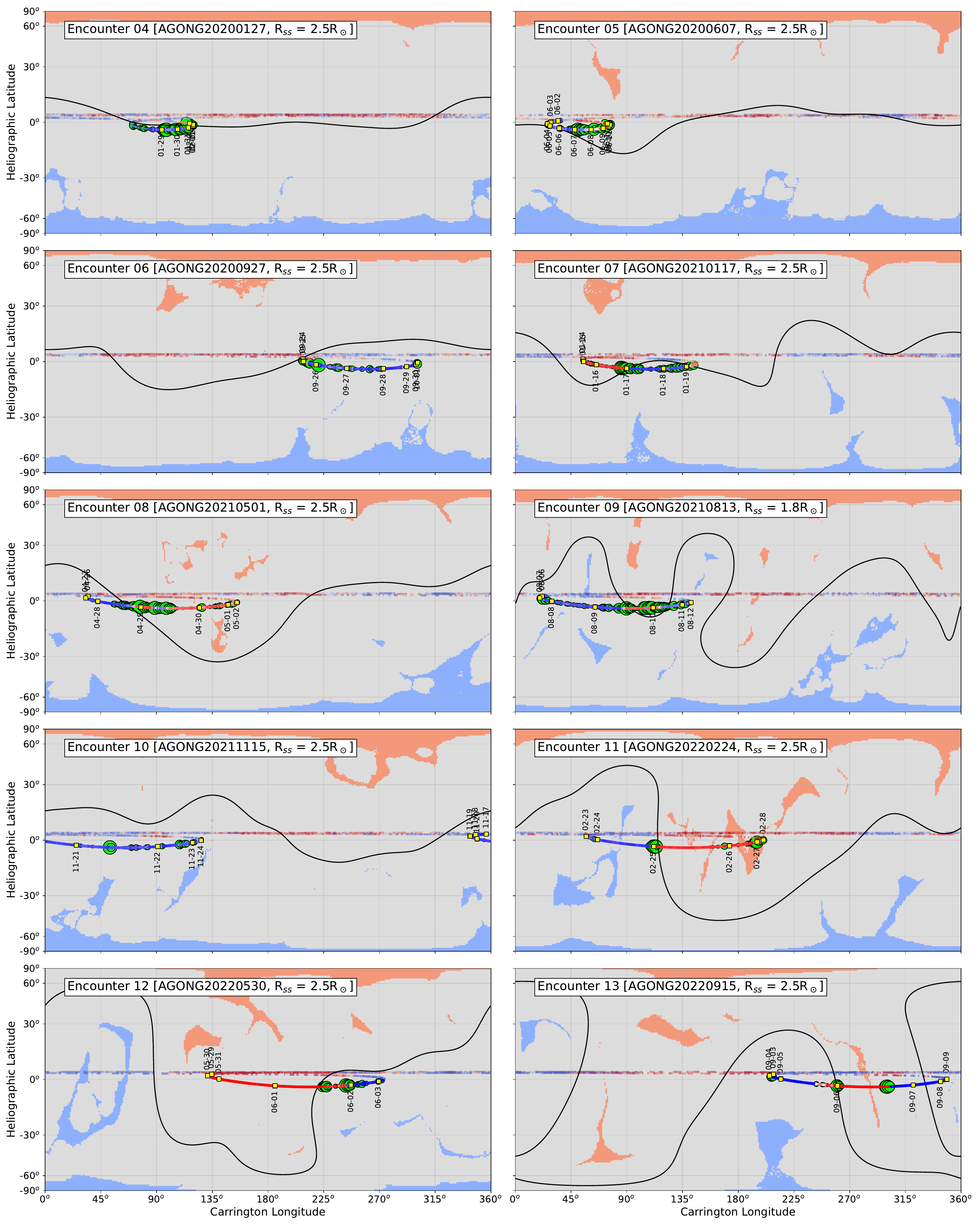}
    \caption{Same as Fig.~\ref{fig:heliographic_maintext} showing (a) PSP trajectory colored by $r^2 B_R(r)$ (\textit{red} indicates positive polarity and \textit{blue} indicates negative polarity), (b) hammerhead occurrence fraction $n_{\mathrm{ham}}/n_{\mathrm{total}}$ proportional to the size of \textit{lime-green} circles centered at each $1^{\circ}$ bin along the PSP trajectory, and (c) the modeled HCS in \textit{solid black} from E04-E13. The \textit{orange} (\textit{blue}) shaded regions indicated coronal holes formed by radially outgoing (incoming) open magnetic field lines.}
    \label{fig:E04-E13}
\end{figure}

\begin{figure}
    \centering
    \includegraphics[width=\linewidth]{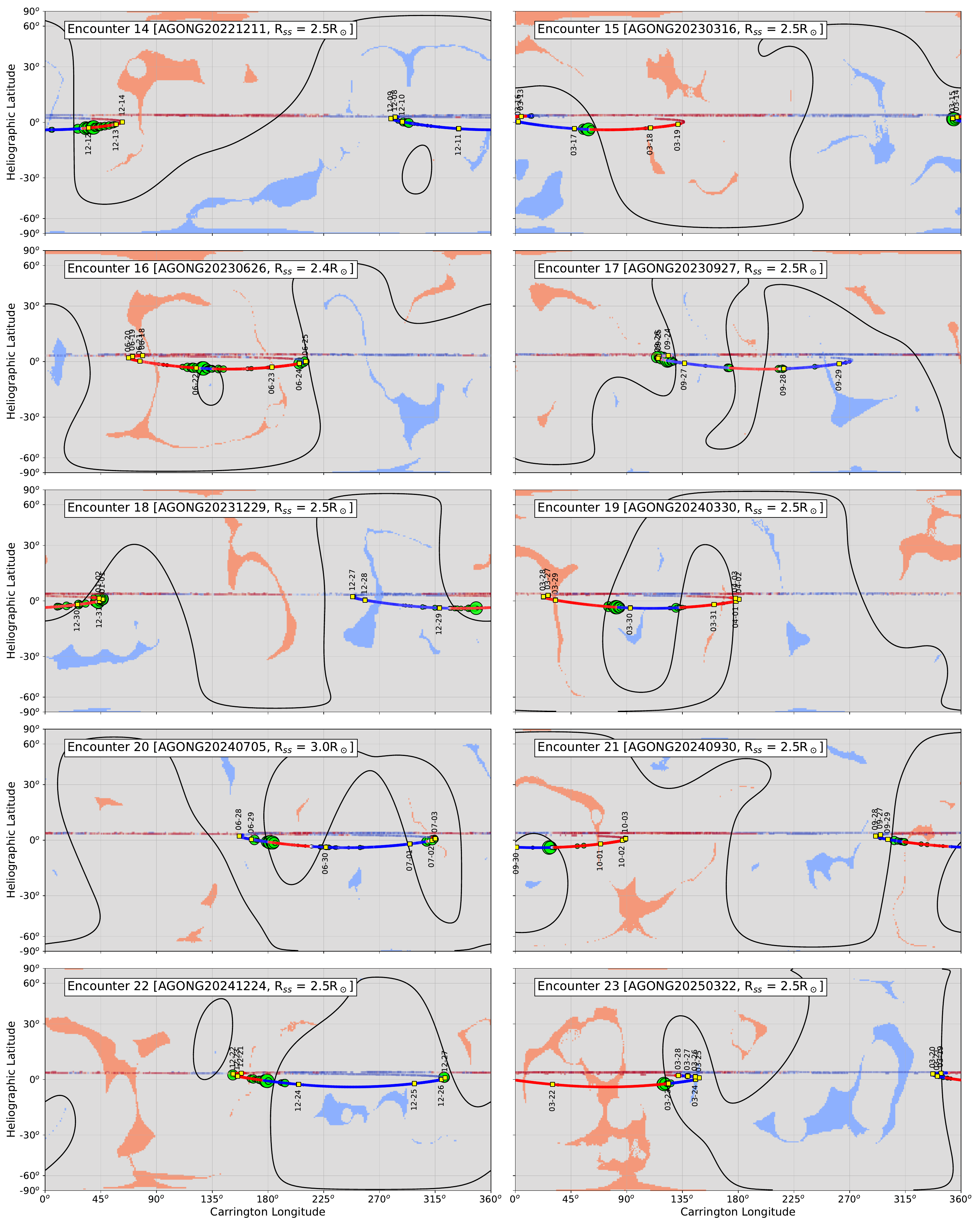}
    \caption{Same as Fig.~\ref{fig:heliographic_maintext} showing (a) PSP trajectory colored by $r^2 B_R(r)$ (\textit{red} indicates positive polarity and \textit{blue} indicates negative polarity), (b) hammerhead occurrence fraction $N_{\mathrm{ham}}/N_{\mathrm{total}}$ proportional to the size of \textit{lime-green} circles centered at each $1^{\circ}$ bin along the PSP trajectory, and (c) the modeled HCS in \textit{solid black} from E14-E23. The \textit{orange} (\textit{blue}) shaded regions indicated coronal holes formed by radially outgoing (incoming) open magnetic field lines.}
    \label{fig:E14-E23}
\end{figure}

We present the PSP trajectory plotted on the heliographic coordinates along an optimized PFSS model for the HCS using \texttt{sunkit-magex} \citep[formerly \texttt{pfsspy,}][]{pfsspy}. 
E04 - E13 is shown in Figure~\ref{fig:E04-E13} and E14 - E23 is shown in Figure~\ref{fig:E14-E23}. 
For each case, the PSP trajectory is colored by the $r^2 \, B_R(r)$ with \textit{red} indicating a positive polarity and \textit{blue} indicating a negative polarity. 
Parts of the trajectory which are not included in the perihelion period where we run \texttt{hampy} are indicated using lower opacity points. 
The \textit{solid black} line indicates the location of the modelled HCS. 
The fractional occurrence of hammerheads $N_{\mathrm{ham}}/N_{\mathrm{total}}$ is proportional to the size of the \textit{lime-green} circles placed at $1^{\circ}$ separation bins on the trajectory. 
Each panel is annotated with the encounter number, the synoptic map used for the magnetogram and the chosen source surface radius $R_{\mathrm{ss}}$.

For completeness, in Table~\ref{tab:pfss}, we provide an accounting of the chosen PFSS input boundary conditions. 
We use ADAPT-GONG \cite{Arge2010} magnetograms, averaged across realizations. 
We take, as a first guess, the date of perihelion for each orbit to select a map and a canonical source surface height of 2.5$R_\odot$ \citep{Hoeksema1984}. 
For cases where a different combination of map and/or surface height has been used in prior work, we defer to these choices. 
Lastly, for instances where there are no prior references and the standard input does not accurately reflect the polarity of the solar wind indicated, we vary the chosen map date and source surface height and visually choose the best matching model. 
It is important to note here that in-situ hammerhead detections can henceforth be used to better guide the PFSS modeling. 
This was done for encounters E09, E12, E16 and E20. 
It is important to note that going forward, given the association between prevalent hammerhead occurrence around HCS crossings established in this work, PFSS tuning even in the absence of an apparent $B_R$ polarity reversal in the in-situ data will be possible. 
Lastly, we provide a comment on any novel hammerhead-HCS geometry connection apparent in a given encounter for future reference.  

\section{Encounter-wise and distance-wise hammerhead properties} \label{sec: Tperp_Tpara_plots}

\begin{figure*}
    \centering
    \includegraphics[width=\linewidth]{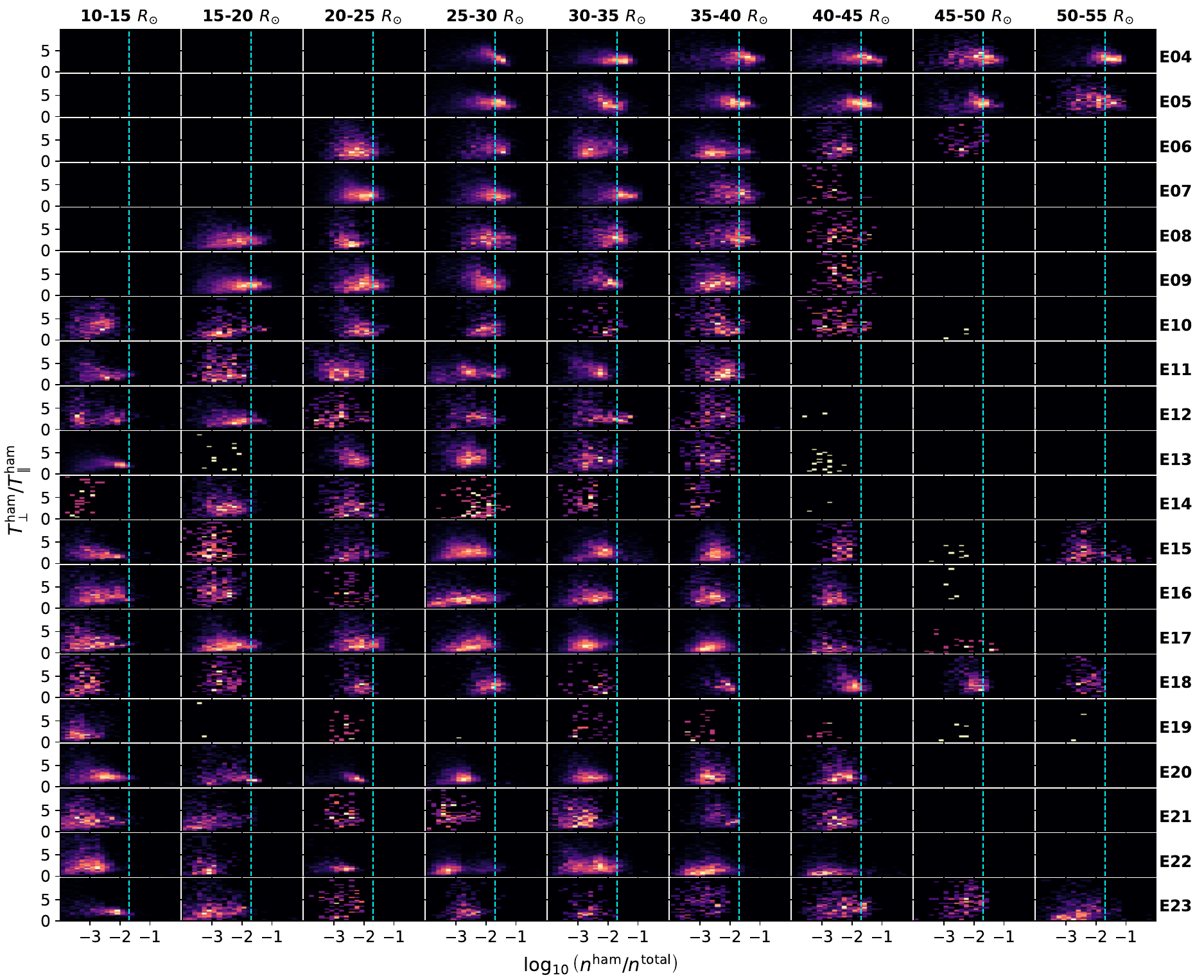}
    \caption{Characterization of hammerhead population detected using \texttt{hampy} from E04 through E23 using trends of $T^{\mathrm{ham}}_{\perp}/T^{\mathrm{ham}}_{\parallel}$ vs. fractional density of \textit{hammer} with respect to total density. The hammerhead detections have been separated into nine radial distance bins for each encounter. Added for visual reference, the \textit{cyan vertical} line indicates the fractional density reported by \cite{Verniero_Beams_2022} who presented the same trends for a 7-hour period on 2020-01-29 during E04. The color-scale is linear and each panel's histogram is normalized to have area under the 2D map to be 1, for better visibility.}
    \label{fig:V22_replot_allEnc_fracden}
\end{figure*}

\begin{figure*}
    \centering
    \includegraphics[width=\linewidth]{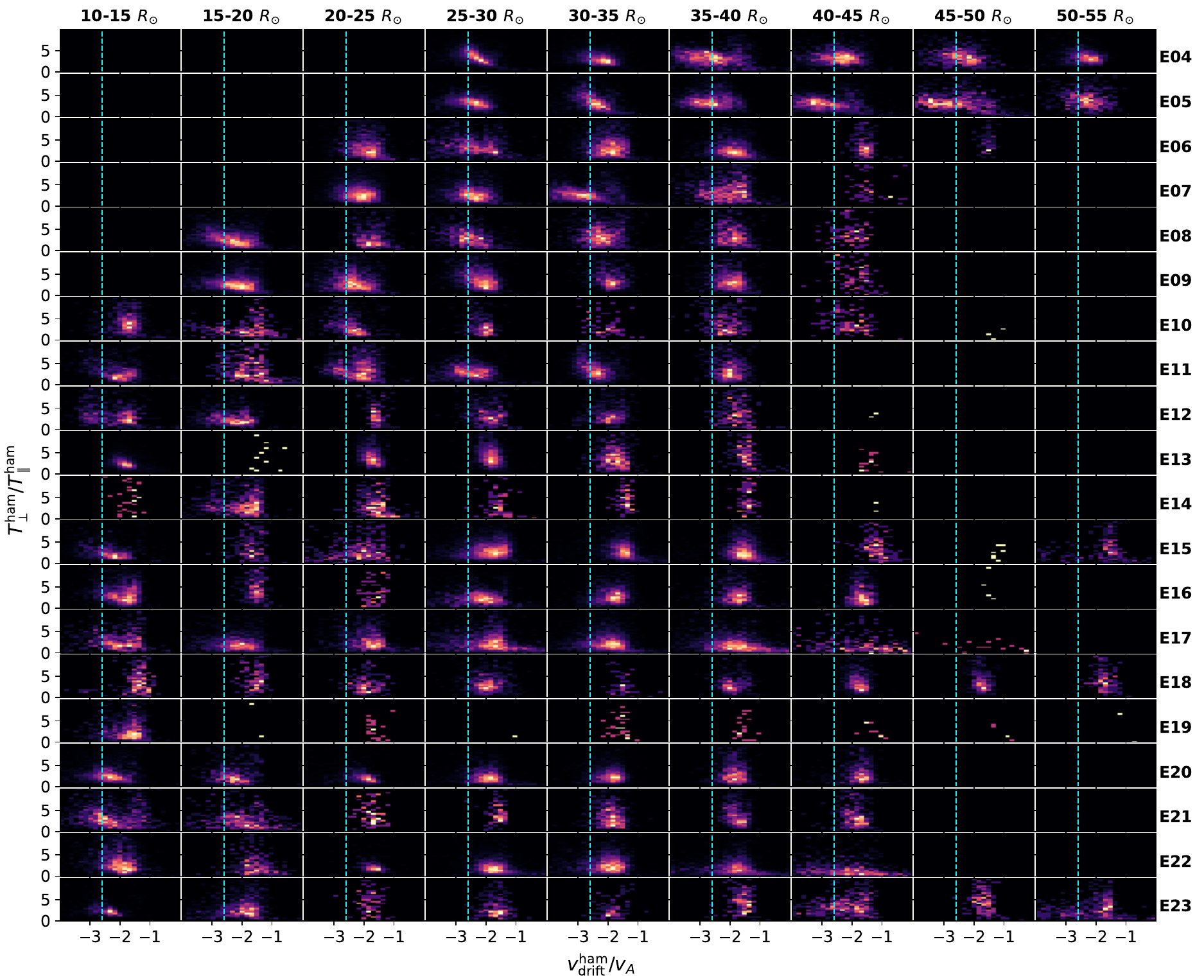}
    \caption{Characterization of hammerhead population detected using \texttt{hampy} from E04 through E23 using trends of $T^{\mathrm{ham}}_{\perp}/T^{\mathrm{ham}}_{\parallel}$ vs. fractional drift velocity of \textit{hammer} with respect to \textit{core} normalized by the Alfv\'en speed $v_{A}$. The hammerhead detections have been separated into nine radial distance bins for each encounter. Added for visual reference, the \textit{cyan vertical} line indicates the fractional density reported by \cite{Verniero_Beams_2022} who presented the same trends for a 7-hour period on 2020-01-29 during E04. The color-scale is linear and each panel's histogram is normalized to have area under the 2D map to be 1, for better visibility.}
    \label{fig:V22_replot_allEnc_fracvel}
\end{figure*}

In this Appendix, we extend the analysis presented in Section~\ref{sec:characterization} across all twenty encounter between radial distances $10R_{\odot}-55 R_{\odot}$ at a bin spacing of $5 R_{\odot}$. 
The panels in Fig.~\ref{fig:V22_replot_allEnc_fracden} show the trends across E04 - E23 for our preliminary investigation of $T^{\mathrm{ham}}_{\perp}/T^{\mathrm{ham}}_{\parallel}$ vs. fractional density of \textit{hammer} with respect to total density. 
Similarly, Fig.~\ref{fig:V22_replot_allEnc_fracvel} show the trends across E04 - E23 for our preliminary investigation of $T^{\mathrm{ham}}_{\perp}/T^{\mathrm{ham}}_{\parallel}$ vs. fractional drift velocity of \textit{hammer} with respect to \textit{core} normalized by the Alfv\'en speed $v_{A}$. 
Although we see, in Sec.~\ref{sec:characterization}, that the \textit{hammer} temperature anisotropy has a clean anti-correlation with both fractional \textit{hammer} density and the scaled drift speed, this is not seen to be consistent across all encounters and all distance range. 
Careful visual inspection of Fig.~\ref{fig:V22_replot_allEnc_fracvel} show panels corresponding to other distance ranges across encounters which demonstrate similar anti-correlation patterns.
However, we do not attempt to draw conclusive inferences regarding these. 
Rather, we defer this task to a future investigation where we will perform detailed case-by-case reconstruction \citep[such as with sophisticated recents methods][]{Das_Terres_ESA} to ensure that partial moments due to FOV restrictions do not bias the trends. 
That said, we do note that there appears to be a weak dependence of the peak of the distribution as a function of encounter. 
For a given radial distance bin, the peak in fractional density migrates to a lower fractional \textit{hammer} density while the peak in scaled drift speed migrates to a lower drift speed. This seems to be connected to the orientation of the HCS with respect to the PSP trajectory where in earlier encounters PSP skim close to a flat HCS and in later encounters PSP cuts across vertical crossings. 
The transition from approximately flat to nearly vertical HCS crossings occur around E09-E11 (see Fig~\ref{fig:E04-E13}). 
This is also where the peak migration is observed to happen in Figs.~\ref{fig:V22_replot_allEnc_fracden} \& \ref{fig:V22_replot_allEnc_fracvel}. 
Future investigations extending this plot to later encounters as the Sun transitions back to a solar minima would reveal the robustness of this apparent solar cycle variation.

% \begin{figure*}
%     \centering
%     \includegraphics[width=\linewidth]{V22_replot_eachEnc.pdf}
%     \caption{Same as Fig.~\ref{fig:V22_replot_allEnc} but plotted for each encounter separately. The total hammerhead count for each encounter is listed at at the top of each panel. The 80\% by-count region is denoted by a white contour.}
%     \label{fig:V22_replot_eachEnc}
% \end{figure*}

% \begin{figure*}
%     \centering
%     \includegraphics[width=0.7\linewidth]{V22_replot_vsDist.pdf}
%     \caption{Same as Fig.~\ref{fig:V22_replot_eachEnc} but binned by distance from the Sun instead of encounters.}
%     \label{fig:V22_replot_vsDist}
% \end{figure*}

\newpage
%\bibliography{references}{}
%\bibliographystyle{aasjournal}

\end{document}